# Intrinsic and extrinsic pinning in NdFeAs(O,F): vortex trapping and lock-in by the layered structure

C. Tarantini,[1*] K. Iida,[2,3] J. Hänisch,[2,4] F. Kurth,[2,5] J. Jaroszynski,[1] N. Sumiya,[3] M. Chihara,[3] T. Hatano,[3] H. Ikuta,[3] S. Schmidt,[6] P. Seidel,[6] B. Holzapfel,[4] D.C. Larbalestier[1]

[1] Applied Superconductivity Center, National High Magnetic Field Laboratory, Florida State University, Tallahassee FL 32310, USA

[2] Institute for Metallic Materials, IFW Dresden, 01171 Dresden, Germany

[3] Department of Crystalline Materials Science, Nagoya University, Chikusa-ku, Nagoya 464-8603, Japan

[4] Institute for Technical Physics, Karlsruhe Institute of Technology, 76344 Eggenstein-Leopoldshafen, Germany

[5] Dresden University of Technology, Faculty for Natural Science and Mathematics, 01062 Dresden, Germany

[6] Friedrich-Schiller-University Jena, Institute of Solid State Physics, 07743 Jena, Germany

The Fe-based superconductors (FBS) present a large variety of compounds whose properties are affected to different extents by their crystal structures. Amongst them, the *RE*FeAs(O,F) (*RE*1111, where *RE* is a rare earth element) is the family with the highest critical temperature $T_c$ but also with a large anisotropy and Josephson vortices as demonstrated in the flux-flow regime in Sm1111 ($T_c \sim$ 55 K). Here we focus our attention on the pinning properties of the lower-$T_c$ Nd1111 in the flux-creep regime. We demonstrate that for $H//c$ critical current density $J_c$ at high temperatures is dominated by point-defect pinning centres, whereas at low temperatures surface pinning by planar defects parallel to the *c*-axis and vortex shearing prevail. When the field approaches the *ab*-planes, two different regimes are observed at low temperatures as a consequence of the transition between 3D-Abrikosov and 2D-Josephson vortices: one is determined by the formation of a vortex staircase structure and one by lock-in of the vortices parallel to the layers. This is the first study on FBS showing this behaviour in a full temperature, field, and angular range and it demonstrates that, despite the lower $T_c$ and anisotropy of Nd1111 with respect to Sm1111, this compound is substantially affected by intrinsic pinning generating a strong *ab*-peak in $J_c$.

[*]e-mail: <u>tarantini@asc.magnet.fsu.edu</u>

Fe-based superconductors (FBS)[1] have been intensely studied in the last years because of the unique physics of their pairing mechanism[2,3] and their potential for applications due to their high critical temperature $T_c$ (up to 55 K)[4,5] and high upper critical field $H_{c2}$ (estimated beyond 100 T).[6,7,8,9] Studies of their critical current density $J_c$ have been also of particular interest because of its weak field dependence and the varieties of properties discovered in the different compounds. BaFe$_2$As$_2$ (Ba122) thin films were widely investigated because of their high density of self-assembled or artificially introduced pinning centres that are able to significantly enhance $J_c$ while also suppressing the effective anisotropy.[10,11,12,13,14] Further pinning improvements were obtained when combining artificially introduced defects with point defects generated by CaF$_2$ substrate-induced strain.[15] No evidence of intrinsic pinning was observed in Ba122; however, enhancements in $J_c(H//ab)$ due to extended correlated planar defects were found.[16] Fe(Se,Te) films, despite their much lower $T_c$, revealed a high $J_c$ [17,18,19] with a low $J_c$ anisotropy[18,19] and surprisingly showed signs of intrinsic pinning below 8 K.[20] The intrinsic pinning properties of SmFeAs(O,F) and LaFeAs(O,F) single crystals were studied by Moll *et al.*[21] by measuring the angular dependence of the vortex dynamics in the flux-flow regime at high current density: they found that a sharp peak appears at low temperature for fields applied parallel to the *ab*-planes. This phenomenon is related to the relatively large intrinsic electronic anisotropy of these materials ($\gamma = (M/m)^{1/2} \sim$ 4-6, where $M$ and $m$ are the effective masses along the *c*-axis and parallel to the *ab*-plane, respectively)[7] that induces a modulation of the order parameter along the *c*-direction. For these anisotropy values the coherence length $\xi_c$ falls below the interlayer distance $d$ with decreasing temperature and the vortices undergo a transformation from 3D Abrikosov to 2D Josephson vortices.[22]

The purpose of this paper is to investigate how the layered structure of NdFeAs(O,F) (Nd1111) influences its vortex dynamics and $J_c$, considering also the multiband effects and the existence of different anisotropy parameters.[23] We studied the pinning properties of a 60 nm-thick Nd1111 thin film deposited on MgO(100) substrate by molecular beam epitaxy (MBE) (see Methods). This film, whose $T_{c,90\%}$ is ~47.1 K ($T_{c,0} \sim$ 42.5 K), was characterized over a wide temperature range and in high magnetic fields up to 35 T at the National High Magnetic Field Laboratory. We found that surface and point defects independently act as pinning centres for *H//c* with two distinct field ranges of effectiveness, whereas for *H//ab* the intrinsic pinning induced by the layered structure of Nd1111 clearly plays a determining role. In this respect we were able to identify *H-T-θ* regions where the vortices are trapped in a staircase structure or locked between the layers.[22]

## Results and analysis

The high crystalline quality of the Nd1111 thin film was verified by X-ray diffraction with narrow full width at half maximum (FWHM) of the 003 rocking curve ($\Delta\omega = 0.62°$) and of the 102 $\phi$-scan ($\Delta\phi = 1.26°$) (see Supplementary and Fig. S1). Field and angular dependence of the critical current density $J_c$ was measured up to 35 T at 4.2 K and up to 16 T at higher temperatures (see Methods). Figure 1 shows how $J_c$ and the pinning force density $F_p$ evolve with temperature for the two main orientations (*H//c* and *H//ab*). At 4.2 K the self-field $J_c$ reaches ~3.3 MA/cm$^2$. The field dependence is very weak in the *ab*-configuration with $J_c$(35T,4.2K) exceeding 1 MA/cm$^2$ and $F_p$(35 T,4.2 K) larger than 400 GN/m$^3$. For *H//c* $J_c$ is clearly more strongly field dependent but still reaches ~$4.8\times10^4$ A/cm$^2$ at 35 T and 4.2 K. Angular dependences were performed up to 35 K (see Supplementary and Fig. S2): in order to identify pinning contributions originating from random uncorrelated defects, correlated defects and intrinsic pinning, the $J_c(\theta)$ curves were analysed using the anisotropic Ginzburg-Landau scaling approach of Blatter *et al.*[24] as shown in Figure 2. According to this, those parts of the $J_c(\theta)$ curves that are affected by small random defects alone collapse onto a single trend line when plotted as a function of the effective field $H_{eff} = H\sqrt{\sin^2\theta + \gamma_{J_c}^{-2}\cos^2\theta}$ (with $\theta$ being the angle between the applied field and the *ab*-planes). To properly rescale the data, an increasing $J_c$ anisotropy parameter $\gamma_{J_c}$ has to be used with decreasing temperature. This behaviour was also observed in the Fe(Se,Te) thin film of ref. 20, which also showed intrinsic pinning. Clear deviations from the main trend approaching the *ab*-direction become more and more obvious at lower temperatures and increasing field (a few examples are marked by black arrows on the 4.2 K data). A much weaker deviation due to correlated defects is also noticed for the data near the *c*-axis at intermediate and high temperatures in the low-field region (few examples are marked by coloured arrows on the 25 K data).

In order to investigate the nature of the different pinning mechanisms in the two principal field configurations, two different approaches have been followed. For the *c*-axis pinning, the shape of the $F_p(H//c)$ curves has been analysed by a modified Dew-Hughes model,[25] but we did not follow a similar approach for $F_p(H//ab)$ because it is unable to reveal a possible 3D/2D transition of the vortices. Moreover, since the Josephson vortices, unlike the Abrikosov vortices, have no normal cores they have little interaction with pinning defects. In order to reveal the nature of the *ab*-pinning, we instead performed an analysis of the *n*-values of the *I-V* curves in the flux-creep regime ($n \sim U_p/k_BT$, where $U_p$ is the pinning potential and $k_B$ is the Boltzmann constant, in the case of logarithmic current density dependence of $U_p$).[22,26]

Fitting $F_p(H//c)$ in Figure 1 with a function $F_p(H) = A(H/H_{Irr})^p(1 - H/H_{Irr})^q$ with 4 free parameters (*p*, *q*, *A* and $H_{Irr}$)[25] generates unphysical *p* and *q* values because of the superposition of different pinning mechanisms. However, using constant *p* and *q* according to different possible pinning scenarios does not reproduce the curves well either, although the best fits were obtained with (*p*,*q*) = (0.5,2) at low temperature and (*p*,*q*) = (1,2) at high temperature. According to ref. 25, these parameters correspond to surface pinning [(0.5,2)] and point defect (*PD*) pinning [(1,2)]. However, vortex shearing generates the same functional dependence as surface pinning and its possible effect has to be considered and will be discussed ahead (surface pinning or vortex shearing contribution will be marked by *S* in the following).[27,28] Considering that this is a thin film, the surface contribution could be provided by planar defects parallel to the *c*-axis such as domain, antiphase/twin boundaries or dislocation array, whereas *PD* pinning is probably induced by atomic defects such as vacancies or disorder. To reproduce the experimental data taking into account the superposition of two contributions we used the expression

$$F_p(H) = F_{p,S}\left(\frac{H}{H_S}\right)^{0.5}\left(1-\frac{H}{H_S}\right)^2 + F_{p,PD}\left(\frac{H}{H_{PD}}\right)\left(1-\frac{H}{H_{PD}}\right)^2 \qquad \text{eq.(1)}$$

with 4 free parameters: $F_{p,S}$, $F_{p,PD}$, $H_S$, $H_{PD}$ [only one contribution is considered for $H > \min(H_S, H_{PD})$]. The first two parameters represent the amplitudes of the *PD* and *S* contributions while the latter two describe the maximum fields of their effectiveness. As shown in Figure 1, this equation well reproduces the $F_p(H//c)$ data at all temperatures. In all cases $H_S > H_{PD}$ implying that $H_S$ actually corresponds to the experimental irreversibility field. The fitting parameters reported in Figure 3(a) reveal a crossover in the dominant mechanism at about 20 K: the *S* contribution dominates at low *T*, whereas *PD* pinning is stronger at high *T*. Figure 3(b) also shows that the *S* contribution has a wider in-field effectiveness with $H_S$ exceeding 50 T at 4.2 K whereas *PD* pinning is limited to ~20 T.

The *I-V* curves, from which $J_c$ was determined, were analysed with the power-law relation $V \sim I^n$ in order to determine the *n*-value that carries information about the pinning potential. In general, in case of random isotropic pinning, *n* scales with $J_c$ independently of temperature, field or angle [in ref. 29 a relation $(n-1) \sim J_c^\alpha$ is suggested, whereas here we found $n \sim J_c^\alpha$]. As a consequence the $J_c(H, \theta)$ and $n(H, \theta)$ plots should have similar trends. This is roughly the case observed in Figure 4 at 30 K (and above, not shown) where *n* presents a maximum along *ab*. However, this trend changes with decreasing temperature. A small dip in *n*-value starts forming at 25 K (it is more visible for increasing field) revealing an inverse *n*-$J_c$ correlation. At 20 K the dip becomes deeper and wider but there is a point exactly at 180° where the *n*-value peaks with respect to its neighbouring angles. At 15 K both the dip and the peak become more marked and visible over a wider angular range. At 10 K, because of the increasing intensity of the emerging peak, the dip substantially disappears leaving just a flattened region at intermediate angles next to the increasingly emergent peak. At 4.2 K there is no longer any obvious trace of the dip and $n(\theta)$ just shows a sharp peak along the *ab*-planes. Even though at any fixed temperature part of the $n(J_c)$ data lie on a single curve showing $n \sim J_c^\alpha$ (Figure 5), clear deviations from this curve occur at higher $J_c$ values, i.e.

when the *ab*-direction is approached. At 25 K, where the dip in $n(\theta)$ starts to appear, the data deviate downward from the $n \sim J_c^\alpha$ trend-line and $J_c$ increases despite the *n*-value drops. This initial behaviour is visible also at 20, 15 and 10 K but the emerging peak inside the dip of $n(\theta)$ produces a second deviation toward high *n*-values. Although less evident, a similar double-deviation behaviour is also observed at 4.2 K.

## Discussion

The critical current density and the pinning properties of the Nd1111 thin film can be compared with other films of the same family. Of particular interest is the comparison with Sm1111 which has the same structure but higher $T_c$. The self-field $J_c$(4.2 K) of this Nd1111 film, ~3.3 MA/cm$^2$, is more than 2.6 times larger than in Sm1111 thin films, despite its lower $T_c$ ($T_{c,90\%}$ ~ 47.1 K in Nd1111 versus 54.2 K in Sm1111).[30] These compounds also show a quite different in-field pinning behaviour. In fact for *H//c*, despite a similar maximum value, $F_p$ of Nd1111 peaks at about 10 T, whereas Sm1111 has a wide plateau between 20 and 35 T. This suggests both a strong difference in the irreversibility field, as expected because of the difference in $T_c$ and in $J_c$(4.2 K,35 T) (3.5 times smaller than in Sm1111), and a different type of active pinning landscape. For *H//ab*, $J_c$(4.2 K,35 T) in the Nd1111 sample is still larger than in the Sm case, suggesting either a larger anisotropy or a more effective pinning mechanism.

Although Blatter rescaling was developed for anisotropic single-band superconductors, it has been successfully applied also to multiband materials like FBS.[20,31,32] However, similarly to what was previously reported for the other compounds, this rescaling reveals a peculiar temperature dependence of the effective $J_c$ anisotropy (inset of Fig. 2): $\gamma_{J_c}$ increases quite sharply with decreasing temperature down to 20 K but it flattens out at lower temperatures. Temperature dependence of anisotropy parameters is typically found in multiband compounds[33] and this was verified also for FBS.[23] In Nd1111 single crystals the $H_{c2}$ anisotropy decreases with decreasing temperature whereas the penetration depth anisotropy $\gamma_\lambda$ increases (the two merging at ~5-7 near $T_c$). In our film $\gamma_{J_c}$ initially increases below $T_c$ similarly to $\gamma_\lambda$, as already reported for Fe(Se,Te) films.[20] However, the magnitude of $\gamma_{J_c}$ is clearly smaller than $\gamma_\lambda$ suggesting that both intrinsic and extrinsic factors affect it. The $J_c$ rescaling also highlights the presence of *ab*- and *c*-axis correlated pinning. The analysis of the $F_p(H//c)$ curves shows the superposition of two independent contributions, *PD* pinning and either surface pinning or vortex shearing. At high temperature the *S* contribution with $(p,q) = (0.5,2)$ can be easily ascribed to surface pinning because of the *c*-axis correlated $J_c$ enhancement revealed by Blatter's rescaling. At lower temperature such enhancement becomes less obvious and vortex shearing may partially determine the *S* contribution. In either case it is interesting to notice that the *PD* and *S* contributions have different field ranges of effectiveness (Figure 3). Although the superposition of different mechanisms is common (e.g. ref. 34,35), their independence as revealed by different effective fields is unusual (one example is given by ref.36 where, however, the high and low field regions were separately fitted using all exponents as free parameters). In our case the lack of interaction can be explained at high temperature by the specific geometry and the type of pinning centres involved. In fact, in an epitaxial thin film the main defects acting as surface pinning are parallel to the *c*-axis. *PD*s are on the other hand typically related to the presence of vacancies or atomic disorder and they are located in the grains. As a consequence, in the *H//c* configuration the vortices are either intragrain and pinned on *PD* or intergrain and pinned on the planar defects parallel to the *c*-axis. Only when the field forms a high angle with the *c*-axis can the vortices interact with both *PD* and planar defects. It is also interesting that the cross-over between the *PD* and *S* contributions occurs at ~20 K [Figure 3(a)], the same temperature at which $\gamma_{J_c}$ changes its trend (inset of Figure 2). *PD*s generate isotropic random pinning that affects $J_c$ at every field angle: the stronger temperature dependence of $\gamma_{J_c}$ in the range dominated by *PD* (>20 K) suggests that there Blatter's rescaling more strongly probes the temperature dependence of the intrinsic anisotropy $\gamma_\lambda$. At lower temperature both $F_{p,PD}$ and $\gamma_{J_c}$ flatten out and the *PD* contribution becomes less important, suggesting that $\gamma_{J_c}$ is affected by factors other than the intrinsic anisotropy $\gamma_\lambda$.

Figures 4 and 5 clearly show that for Nd1111 the $n \sim J_c^{\alpha}$ relation does not hold in the whole $T$-$H$-$\theta$ range and that two drastic changes in the $n$-$J_c$ trend indicate two different pinning regimes. The $n(J_c)$ plots are particularly helpful in determining the angles from the $ab$-plane at which these deviations occur and these transition angles are reported in Figure 6. The suppression of the $n$-value, which starts to be visible at 25 K, affects an angular region up to an angle $\varphi_T$ (trapping angle) either side of the $ab$-plane. This behaviour was previously observed in high-$T_c$ superconductors like YBa$_2$Cu$_3$O$_{7-\delta}$ (YBCO)[37,38] and more recently in other FBS like Sm1111[30] and Fe(Se,Te)[20] where it has been ascribed to the formation of a staircase structure in which segments of vortices are trapped in the $ab$-plane and connected by vortex kinks as sketched in Figure 6(a). With the current parallel to the planes and perpendicular to $H$, the Lorentz force on the vortices is directed along the $c$-axis for the portion trapped between the $ab$-planes but it is mostly along the $ab$-planes for the kinked segments.[39] This induces motions of the kinks and suppression of the $n$-value. Up to 20 K, the field dependence of $\varphi_T$ approximately follows the $\varphi_T \sim H^{-3/4}$ relation [Figure 6(a)] as theoretically predicted by Blatter $et$ $al$.,[22] whereas at 25 K it has a weaker field dependence probably due to the proximity to the 2D/3D transition. The 2D/3D transition occurs at a temperature $T_{cr}$ between 25 and 30 K: knowing the layer spacing in the Nd1111 sample ($d = 0.856$ nm determined by XRD), $\xi_c(0)$ can be calculated by the relation $\xi_c(0) = d\sqrt{(1-T_{cr}/T_c)/2}$ (ref. 22) to be 0.36-0.41 nm. Since the in-plane coherence length $\xi_{ab}(0)$ (estimated from the $H_{c2}$ slope at $T_c$ and the WHH formula[40]) is about 1.93 nm, the intrinsic anisotropy $\gamma = \xi_{ab} / \xi_c$ lies between 4.7 and 5.4. These anisotropy and coherence length values are consistent with those previously reported.[7,41] It is important to notice however that the relations in ref. 22 assume a temperature independent anisotropy but, since this is not the case for FBS,[23] the $\gamma$ value estimated here should be ascribed to the temperature range for which it was calculated (25-30 K). It is also interesting to notice that $T_{cr}/T_c$ is about 0.6, indicating that Nd1111 is an intermediate case with respect to La1111 ($T_{cr}/T_c \sim 0.5$) and Sm1111 ($T_{cr}/T_c \sim 0.8$)[21] as already suggested from $H_{c2}$ characterizations.[7]

Figure 6(b) shows the field dependence of $\varphi_L$ (lock-in angle, $\varphi_L < \varphi_T$): this is the maximum angle from the *ab*-plane at which the vortices are completely locked parallel to the layers, as sketched in Figure 6(b). Since in this case the whole vortex is parallel to the *ab*-planes, the Lorentz force is always directed along the *c*-axis. This generates very strong pinning and an increased *n*-value (Figure 4 and 5). $\varphi_L$ seems to have a weaker dependence ($\sim H^{1/2}$) than the theoretical prediction ($\varphi_L \sim \varphi_T / H$)[22] and an amplitude larger than previously observed in YBCO (0.1-1°)[22,39] which has a slightly higher intrinsic anisotropy. However, here the reduced temperature is much lower than for YBCO and the multiband nature of FBS and the temperature dependence of the superconducting parameters could also play a role in determining the vortex lock-in. Another factor to take into account is the sample mosaicity: despite the high crystalline quality, the vortex lock-in likely occurs over the domain-size, not the entire sample, enhancing $\varphi_L$. Awaji *et al.*[38] characterized YBCO films down to 4.2 K and observed that the *n*-value for *H//ab* first increases on going below $T_c$, has a plateau from 70 to 40 K and then increases again below 20 K. The authors explained this behaviour by vortex kink excitation in the plateau region, followed by its suppression below 20 K. In our Nd1111 sample we observed the same trend when the reduced temperature is taken into account. Clear evidence of locked-in vortices have never been reported before for FBS, however Iida *et al.*[20] did observe a small peak emerging from the *n*-value dip at 4 K in Fe(Se,Te). It is quite striking that even a material with such a low $T_c$ reveals itself to have such strong intrinsic pinning.

To conclude, in this paper we investigated the intrinsic and extrinsic pinning properties of an epitaxial NdFeAs(O,F) thin film by measuring the field and angular dependence of $J_c$ up to 35 T. We found that both intrinsic and extrinsic pinning contributions affect the $J_c$ performance. $J_c(H//c)$ is dominated by extrinsic pinning properties determined by the sample microstructure and $F_p$ can be described at every temperature by the superposition of surface and point pinning contributions or vortex shearing acting in different field ranges. More striking is the clear evidence of intrinsic pinning below 25-30 K induced by the layered structure of Nd1111 when the *ab*-planes are approached. In fact, when the magnetic field forms an angle smaller than $\varphi_T$ from the *ab*-plane for *T* < 30 K, a staircase structure of vortices is created that produces an inverse correlation between $J_c$ and $n$ because of the movement of kinked vortex segments. At lower temperature ($T \leq 20$ K) and angles smaller than $\varphi_L$ ($<\varphi_T$), the vortices are locked between the strong pinning layers, greatly increasing the *n*-value. These results make clear that the 1111 phase has intrinsic properties more similar to the high-$T_c$ superconductors like YBCO than to low-$T_c$ materials, an important factor to be taken into account when considering this compound for possible applications.

**Methods**

NdFeAs(O,F) epitaxial thin films have been prepared on MgO(100) single crystal substrates by molecular beam epitaxy (MBE) using solid sources of $NdF_3$, Fe, As and Ga and a gas source of $O_2$. Here Ga works as F getter as the following reaction occurs: Ga + 3F → $GaF_3$. A first deposition at 800 °C yielded the mother compound of NdFeAsO with a thickness of about 75 nm. Subsequently, the NdOF cap layer was deposited on the NdFeAsO at 800 °C, followed by annealing at the same temperature for 0.5 h. During the NdOF cap layer deposition and annealing processes, F diffuses into the NdFeAsO layer. The detailed fabrication process was reported in ref. 42. In order to remove the NdOF caplayer, the resultant film was covered by photolithography, followed by ion-beam etching. After the etching process the NdFeAs(O,F) film thickness was about 60 nm.

Transport characterizations were performed on 45 μm wide and 1 mm long bridges fabricated by laser cutting. The *I-V* curves were obtained at varying field, temperature and field orientation but always while maintaining the maximum Lorentz force configuration. Data were measured up to 16 T in a physical property measurement system (PPMS) and up to 35 T in the 35 T DC magnet at the National High Magnetic Field Laboratory (NHMFL) in Tallahassee. The critical current was determined by a 1 μV/cm criterion.

**References**

[1] Kamihara, Y., Watanabe, T., Hirano, M. & Hosono, H. Iron-Based Layered Superconductor La[$O_{1-x}F_x$]FeAs (x = 0.05–0.12) with $T_c$ = 26 K. *J. Am. Chem. Soc*. 130, 3296–3297 (2008).

[2] Paglione, J. & Greene, R. L. High-temperature superconductivity in iron-based materials. *Nature Phys.* 6, 645 (2010).

[3] Johnston, D. C. The puzzle of high temperature superconductivity in layered iron pnictides and chalcogenides. *Adv. Phys*. 59, 803 (2010).

[4] Ren, Z. A. *et al.* Superconductivity at 55 K in Iron-Based F-Doped Layered Quaternary Compound Sm[O12xFx]FeAs. *Chinese Phys. Lett.* 25, 2215–2216 (2008).

[5] Rotter, M., Tegel, M. & Johrendt, D. Superconductivity at 38 K in the Iron Arsenide ($Ba_{1-x}K_x$)$Fe_2As_2$. *Phys. Rev. Lett*. 101, 107006 (2008).

[6] Hunte, F., *et al.* Two-band superconductivity in LaFeAsO$_{0.89}$F$_{0.11}$ at very high magnetic fields, *Nature* 453, 903 (2008).

[7] Jaroszynski, J. *et al.* Comparative high-field magnetotransport of the oxypnictide superconductors RFeAsO$_{1-x}$F$_x$ (R=La, Nd) and SmFeAsO$_{1-\delta}$. *Phys. Rev. B* 78, 064511 (2009).

[8] Lee, H. *et al*. Effects of two gaps and paramagnetic pair breaking on the upper critical field of SmFeAsO0.85 and SmFeAsO$_{0.8}$F$_{0.2}$ single crystals. *Phys. Rev. B* 80, 144512 (2009).

[9] Tarantini, C. *et al*. Significant enhancement of upper critical fields by doping and strain in iron-based superconductors. *Phys. Rev. B* 84, 184522 (2011).

[10] Tarantini, C. *et al.* Strong vortex pinning in co-doped BaFe$_2$As$_2$ single crystal thin films. *Appl. Phys. Lett*. 96, 142510 (2010).

[11] Maiorov, B. *et al*. Liquid vortex phase and strong *c*-axis pinning in low anisotropy BaCo$_x$Fe$_{2-x}$As$_2$ pnictide films, *Supercond. Sci. Technol*. 24, 055007 (2011).

[12] Tarantini, C. *et al*. Artificial and self-assembled vortex-pinning centers in superconducting Ba(Fe$_{1-x}$Co$_x$)$_2$As$_2$ thin films as a route to obtaining very high critical-current densities. *Phys. Rev. B* 86, 214504 (2012).

[13] Miura, M. *et al*. Strongly enhanced flux pinning in one-step deposition of BaFe$_2$(As$_{0.66}$P$_{0.33}$)$_2$ superconductor films with uniformly dispersed BaZrO$_3$ nanoparticles. *Nat. Comm.* 4, 2499 (2013).

[14] Sato, H., Hiramatsu, H., Kamiya, T. & Hosono, H. High critical-current density with less anisotropy in BaFe$_2$(As,P)$_2$ epitaxial thin films: Effect of intentionally grown *c*-axis vortex-pinning centers. *Appl. Phys. Lett.* 104, 182603 (2014).

[15] Tarantini, C. *et al*. Development of very high $J_c$ in Ba(Fe$_{1-x}$Co$_x$)$_2$As$_2$ thin films grown on CaF$_2$. *Sci. Rep.* 4, 7395 (2014).

[16] Hänisch, J. *et al.* High field superconducting properties of Ba(Fe$_{1-x}$Co$_x$)$_2$ As$_2$ thin films. *Sci. Rep.* 5, 17363 (2015).

[17] Iida, K. *et al.* Generic Fe buffer layers for Fe-based superconductors: Epitaxial FeSe$_{1-x}$Te$_x$ thin films. *Appl. Phys. Lett.* 99, 202503 (2011).

[18] Bellingeri, E. *et al.* Strong vortex pinning in FeSe$_{0.5}$Te$_{0.5}$ epitaxial thin film. *Appl. Phys. Lett.* 100, 082601 (2012).

[19] Si, W. *et al*. High current superconductivity in FeSe0.5Te0.5-coated conductors at 30 tesla.. *Nat. Commun*. 4, 1347 (2013).

[20] Iida, K. *et al*. Intrinsic pinning and the critical current scaling of clean epitaxial Fe(Se,Te) thin films. *Phys. Rev. B* 87, 104510 (2013).

[21] Moll, P.J.W. *et al*. Transition from slow Abrikosov to fast moving Josephson vortices in iron pnictide superconductors, *Nature Mat.*12,134 (2013).

[22] Blatter, G., Feilgl'man, M., Geshkenbin, V. B., Larkin, A. I. and Vinokur, V. Vortices in high-temperature superconductors. *Rev. Mod. Phys*. 66, 1125 (1994).

[23] Weyeneth, S. *et al.* Evidence for two distinct anisotropies in the oxypnictide superconductors SmFeAsO$_{0.8}$F$_{0.2}$ and NdFeAsO$_{0.8}$F$_{0.2}$. *J. Supercond. Nov. Magn.* 22, 347 (2009).

[24] Blatter, G., Geshkenbein, V.B., Larkin, A.I. From isotropic to anisotropic superconductors: a scaling approach. *Phys. Rev. Lett.* 68, 875 (1992).

[25] Dew-Hughes, D. Flux pinning mechanisms in type II superconductors, *Philos. Mag.* 30, 293 (1974).

[26] Zeldov, E. *et al.* Flux creep characteristics in high–temperature superconductors. *Appl. Phys. Lett.* 56, 680–682 (1990).

[27] Kramer, E.J. *et al.* Scaling laws for flux pinning in hard superconductors. *J. Appl. Phys.* 44, 1360 (1973).

[28] Dew-Hughes, D. Is J$_c$ in Nb$_3$Sn limited by grain boundary flux-shear? *IEEE Trans. Magn.* 23, 1172 (1987).

[29] Taylor, D.M.J. and Hampshire, D.P. Relationship between the n-value and critical current in Nb$_3$Sn superconducting wires exhibiting intrinsic and extrinsic behavior. *Supercond. Sci. Technol.* 18, 297 (2005).

[30] Iida, K. *et al.* Oxypnictide SmFeAs(O,F) superconductor: a candidate for high-field magnet applications. *Sci. Rep.* 3, 2139 (2013).

[31] Kidszun, M. *et al.* Critical Current Scaling and Anisotropy in Oxypnictide Superconductors. *Phy. Rev. Lett.* 106, 137001 (2011).

[32] Iida, K. *et al.* Scaling behavior of the critical current in clean epitaxial Ba(Fe$_{1-x}$Co$_x$)$_2$As$_2$ thin films. *Phys. Rev. B* 81, 100507(R) (2010).

[33] Fletcher, J.D., Carrington, A., Taylor, O.J., Kazakov, S.M. and Karpinski, J., Temperature-Dependent Anisotropy of the Penetration Depth and Coherence Length of MgB$_2$. *Phys. Rev. Lett.* 95, 097005 (2005).

[34] Baumgartner, T. *et al.* Effects of neutron irradiation on pinning force scaling in state-of-the-art Nb$_3$Sn wires. *Supercond. Sci. Technol.* 27, 015005 (2014).

[35] Leo, A. *et al.* Vortex pinning properties in Fe-chalcogenides. *Supercond. Sci. Technol*. 28, 125001 (2015).

[36] Li, P.J. *et al.* The collective flux pinning force in $(Sm_{0.5}Eu_{0.5})Ba_2Cu_3O_{7-\delta}$ superconductors. *Supercond. Sci. Technol*. 19, 825, (2006).

[37] Civale, L. *et al.* Identification of intrinsic ab-plane pinning in $YBa_2Cu_3O_7$ thin films and coated conductors. *IEEE Trans. Appl. Supercond.,* 15, 2808 (2005).

[38] Awaji, S. *et al.* Anisotropy of the critical current density and intrinsic pinning behaviors of $YBa_2Cu_3O_y$ coated conductors. *Appl. Phys. Express* 4, 013101 (2011).

[39] Kwok, W. K. *et al*. Direct observation of intrinsic pinning by layered structure in single-crystal $YBa_2Cu_3O_{7-\delta}$. *Phys. Rev. Lett.* 67, 390 (1991).

[40] Werthamer, N. R., Helfard, E. and Hohenberg, P. C. Temperature and purity dependence of the superconducting critical field, $H_{c2}$. III. Electron spin and spin-orbit effects. *Phys. Rev*. 147, 295 (1966).

[41] Jia, Y. *et al.* Critical fields and anisotropy of $NdFeAsO_{0.82}F_{0.18}$ single crystals. *Appl. Phys. Lett.* 93, 032503 (2008).

[42] Uemura, H. *et al*. Substrate dependence of the superconducting properties of NdFeAs(O,F) thin films. *Solid State Commun*. 152, 735 (2012).

**Acknowledgements**

A portion of this work was performed at the National High Magnetic Field Laboratory, which is supported by National Science Foundation Cooperative Agreement No. DMR-1157490 and the State of Florida. The research leading to these results has received funding from European Union's Seventh Framework Programme (FP7/2007-2013) under grant agreement number 283141 (IRON-SEA). This research has been also supported by Strategic International Collaborative Research Program (SICORP), Japan Science and Technology Agency.

**Author Contributions**

C.T., K.I. and J.H. designed the study. C.T., K.I., F.K. and J.J. carried out the high field measurements. C.T. performed the low field characterization and prepared the manuscript with K.I., J.H., as well as D.C.L and B.H. Thin film preparation, structural characterizations, and surface treatment were conducted by F.K., S.S., M.C., N.S., T.H., P.S. and H.I. All authors discussed the results and implications and commented on the manuscript.

**Additional information**

Competing financial interests: The authors declare no competing financial interests.

**Figure Legends**

**Figure 1 Field dependence of critical current density $J_c$ and pinning force density $F_p$ of a NdFeAs(O,F) thin film.** The film was measured with field parallel to the *c*-axis (a,b) and the *ab*-plane (c,d) up to 16 T in the 35-10 K temperature range and in high-field up to 35 T at 4.2 K. The red lines in panel (b) are fitting curves obtained with eq. (1) as described in the text.

**Figure 2 Blatter's rescaling of the angular dependence $J_c(\theta)$.** The curves were measured up 16 T in the 35-10 K range and up to 35 T at 4.2 K. Black arrows along the 4.2 K data indicate deviations from the rescaling due to *ab*-correlated pinning, whereas coloured arrows along the 25 K data point to deviations related to *c*-axis correlated pinning. Inset: temperature dependence of the anisotropy $\gamma_{J_c}$ as obtained from the rescaling.

**Figure 3 Surface and point defects pinning contributions to $F_p(H//c)$.** Fitting parameters of the curves in Figure 1(b) according to eq. (1) showing the temperature dependence of (a) the *S* and *PD* amplitudes and (b) their effectiveness field range.

**Figure 4 Angular dependence of the *n*-value in the 30-4.2 K temperature range.** Magnetic field was applied up to 16 T in the 30-10 K temperature range and in high-field up to 35 T at 4.2 K.

**Figure 5 *n*-value as a function of $J_c$ in the 30-4.2 K temperature range.** Magnetic field was applied up to 16 T in the 30-10 K temperature range and in high-field up to 35 T at 4.2 K. The black lines are linear eye-guides and the red arrows emphasize the trend followed by the data approaching the *ab*-direction.

**Figure 6 Field dependence of the trapping, $\varphi_T$, and lock-in , $\varphi_L$, angles.** The angles were obtained from figures 4 and 5. The insets represent the sketches of the arrangement of vortices (red lines) in the trapping (a) and lock-in (b) regimes in a layered superconductor.

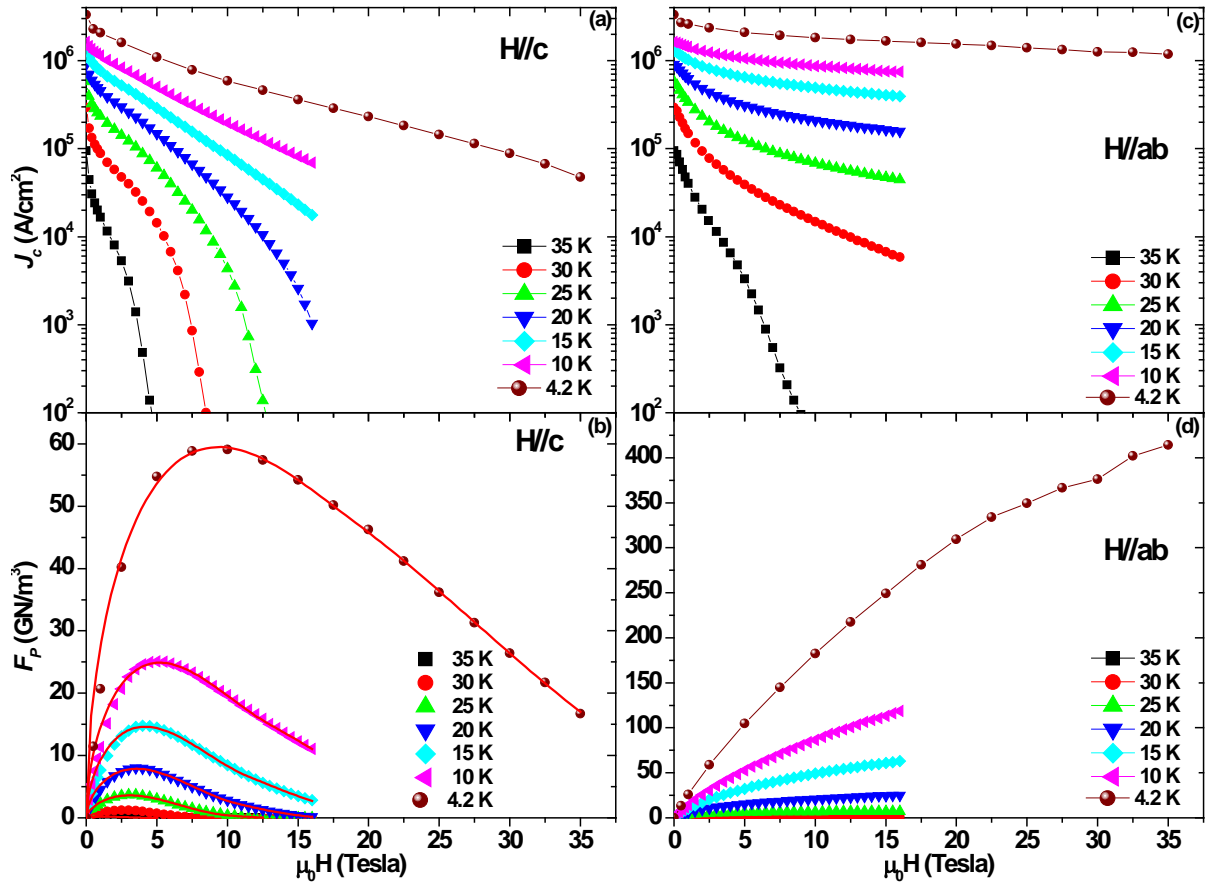

Figure 1

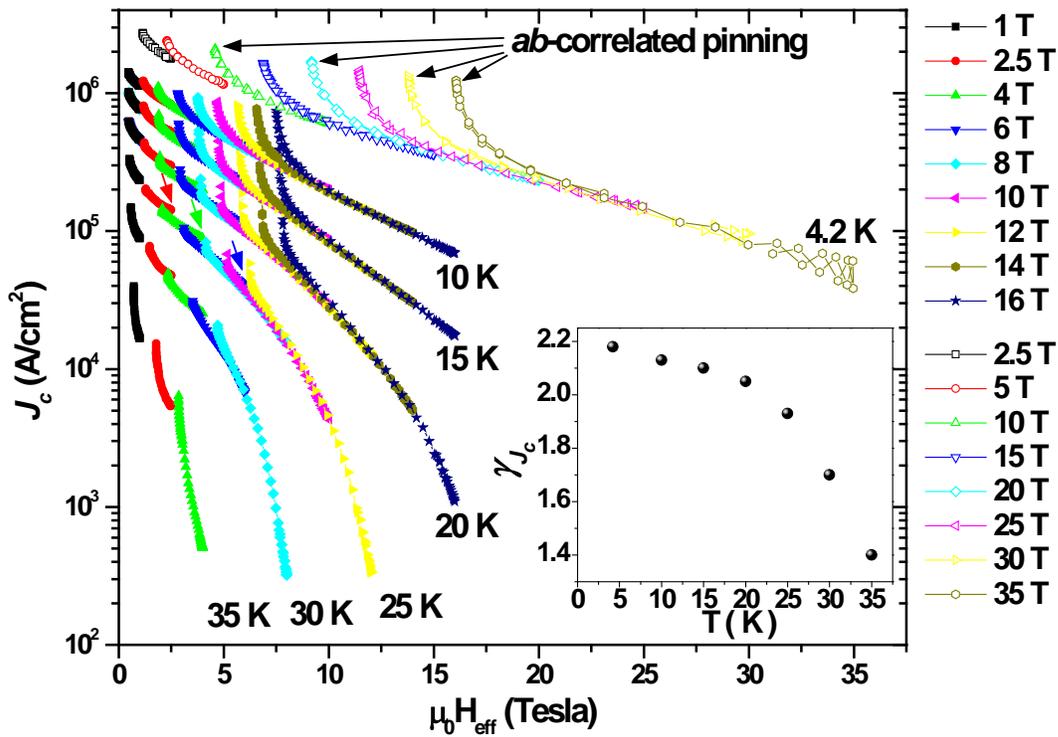

Figure 2

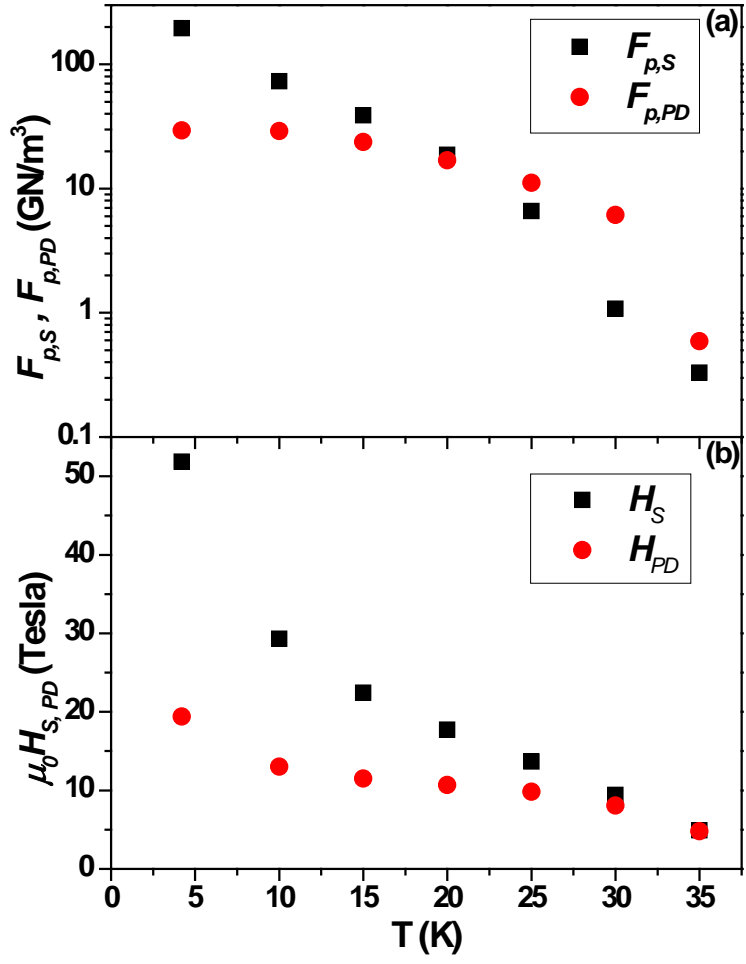

**Figure 3**

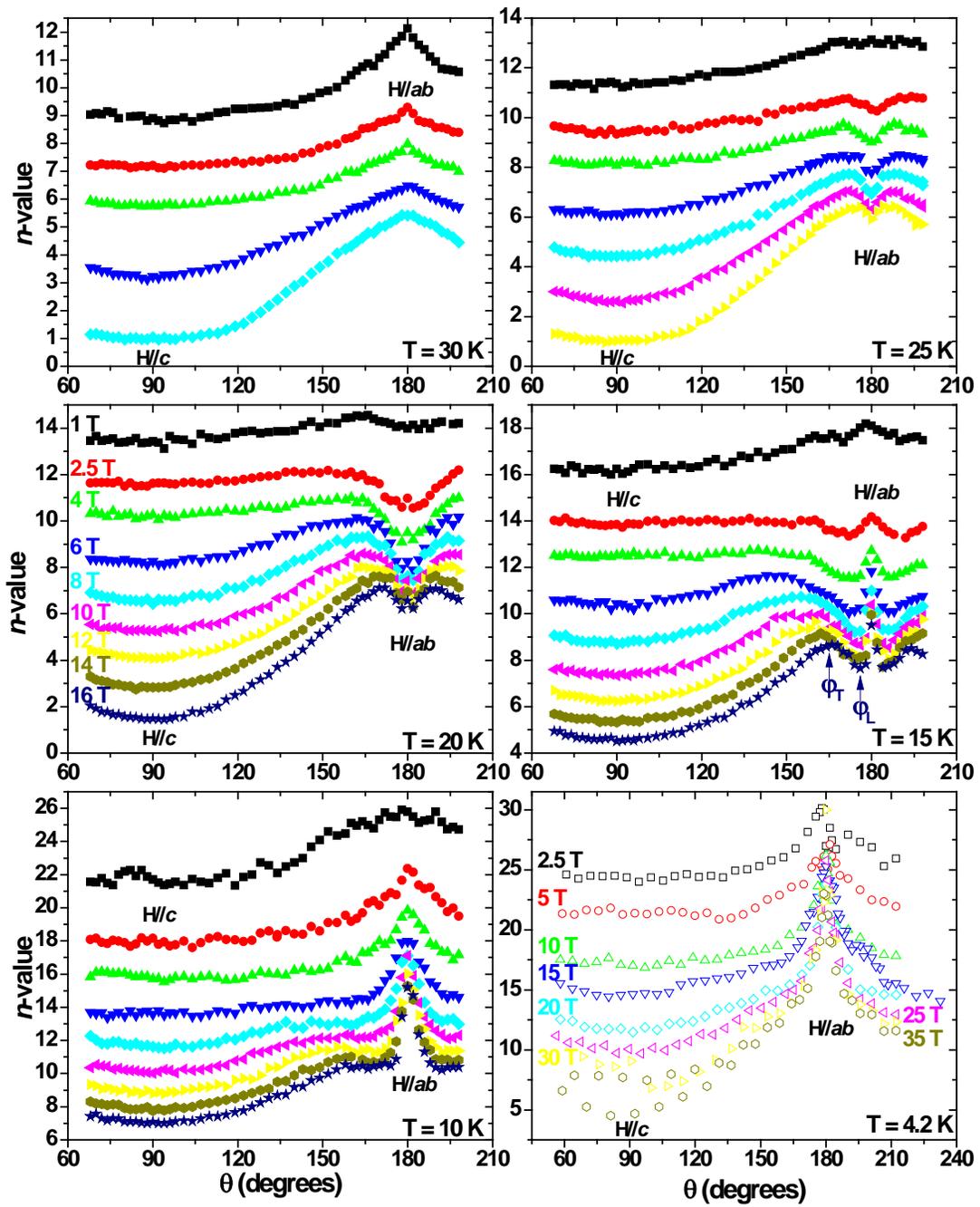

Figure 4

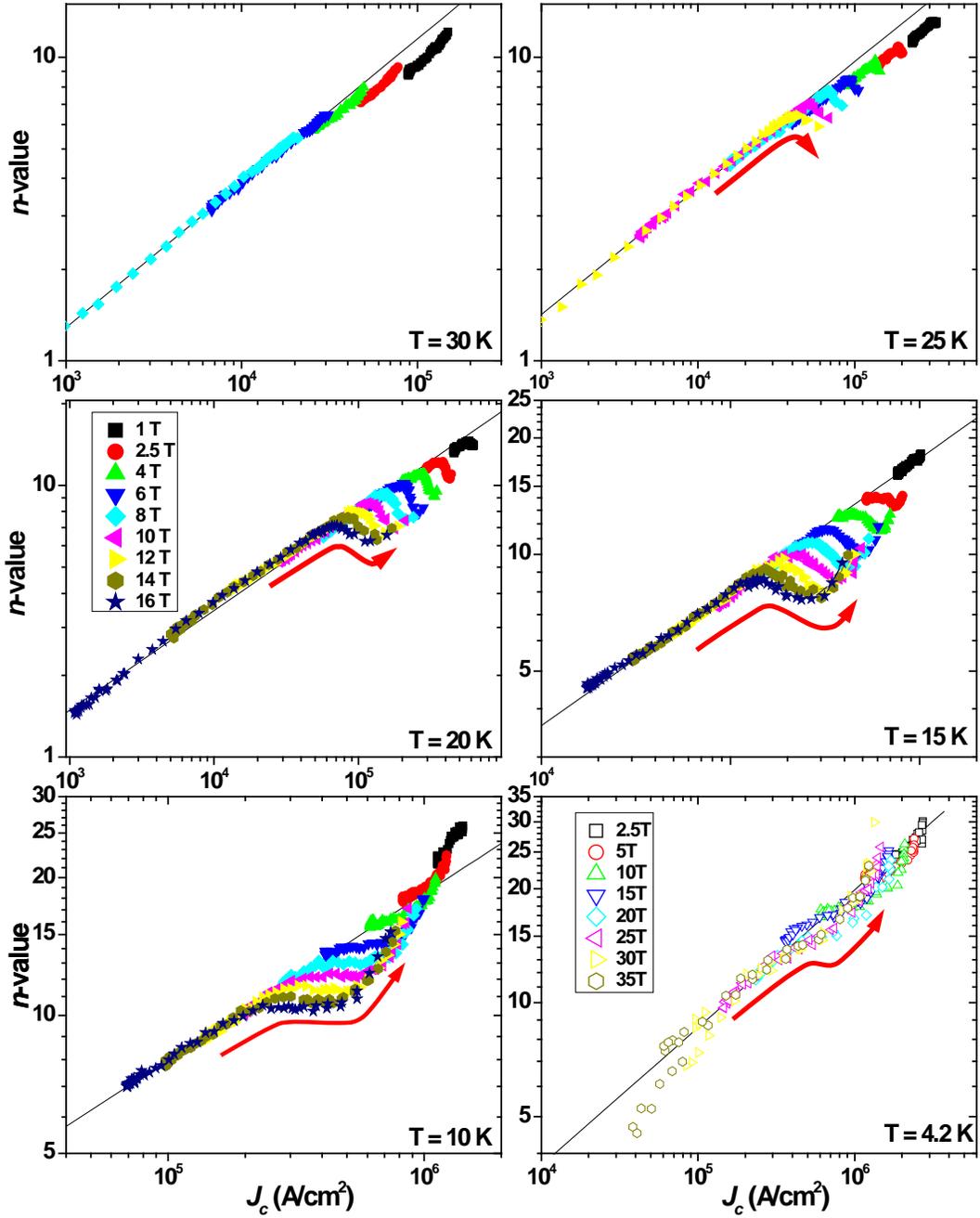

**Figure 5**

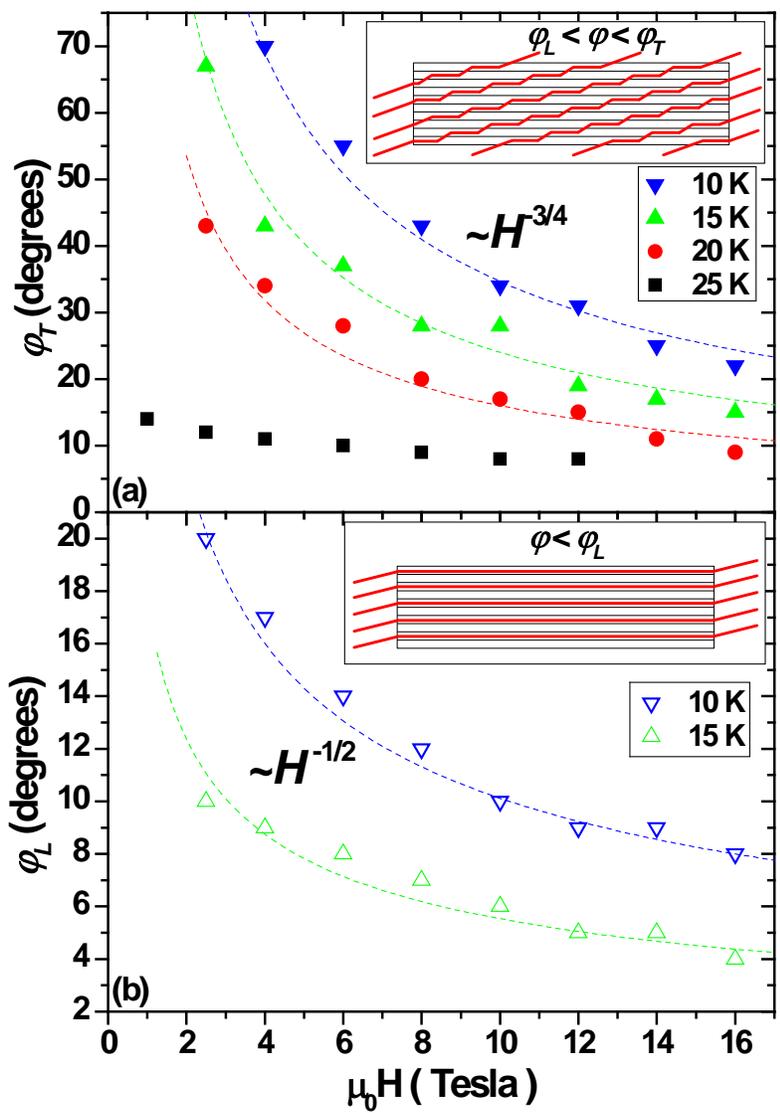

**Figure 6**

# SUPPLEMENTARY INFORMATION FOR: Intrinsic and extrinsic pinning in NdFeAs(O,F): observation of trapped and locked-in vortices by layered structure

C. Tarantini,[1*] K. Iida,[2,3] J. Hänisch,[2,4] F. Kurth,[2,5] J. Jaroszynski,[1] N. Sumiya,[3] M. Chihara,[3] T. Hatano,[3] H. Ikuta,[3] S. Schmidt,[6] P. Seidel,[6] B. Holzapfel,[4] D.C. Larbalestier[1]

[1] Applied Superconductivity Center, National High Magnetic Field Laboratory, Florida State University, Tallahassee FL 32310, USA

[2] Institute for Metallic Materials, IFW Dresden, 01171 Dresden, Germany

[3] Department of Crystalline Materials Science, Nagoya University, Chikusa-ku, Nagoya 464-8603, Japan

[4] Institute for Technical Physics, Karlsruhe Institute of Technology, 76344 Eggenstein-Leopoldshafen, Germany

[5] Dresden University of Technology, Faculty for Natural Science and Mathematics, 01062 Dresden, Germany

[6] Friedrich-Schiller-University Jena, Institute of Solid State Physics, 07743 Jena, Germany

The NdFeAs(O,F) thin film was characterized by XRD. The $\theta$-$2\theta$ scan in Fig. S1(a) shows only the (00$l$) reflections of the NdFeAs(O,F) phase. The (003) rocking curve has a narrow full width at half maximum (FWHM) of $\Delta\omega = 0.62°$ [Fig. S1(b)]. Both demonstrate an excellent out-of-plane orientation. The (102) $\phi$-scan of NdFeAs(O,F) in Fig. S1(c) exhibits a sharp FWHM, $\Delta\phi = 1.26°$, and reveals a fourfold symmetry indicative of epitaxial growth with a (001)[100]NdFeAs(O,F)//(001)[100]MgO relation (cube-on-cube).

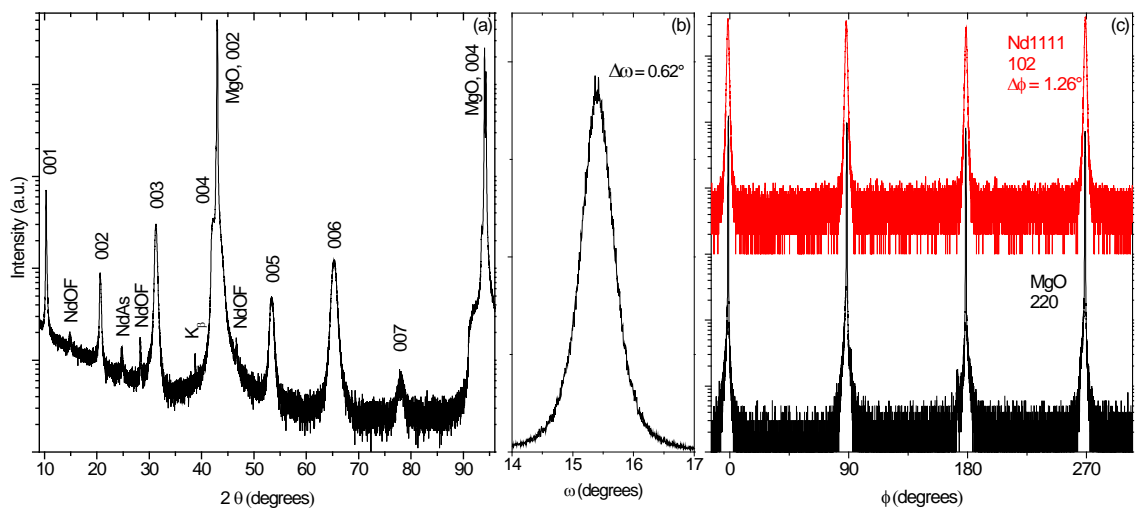

**Figure S1** (a) $\theta$-$2\theta$ scan of NdFeAs(O,F) measured after the ion beam etching. (b) Rocking curve of the (003) reflection of NdFeAs(O,F). (c) The $\phi$-scans on the (102) peak of NdFeAs(O,F) thin film and on the (220) peak of MgO substrate.

The angular dependencies of $J_c$ were measured in a wide 35-4.2 K temperature range up to 35 T at 4.2 K and up to 16 T at higher temperatures. The data are reported in Fig. S2.

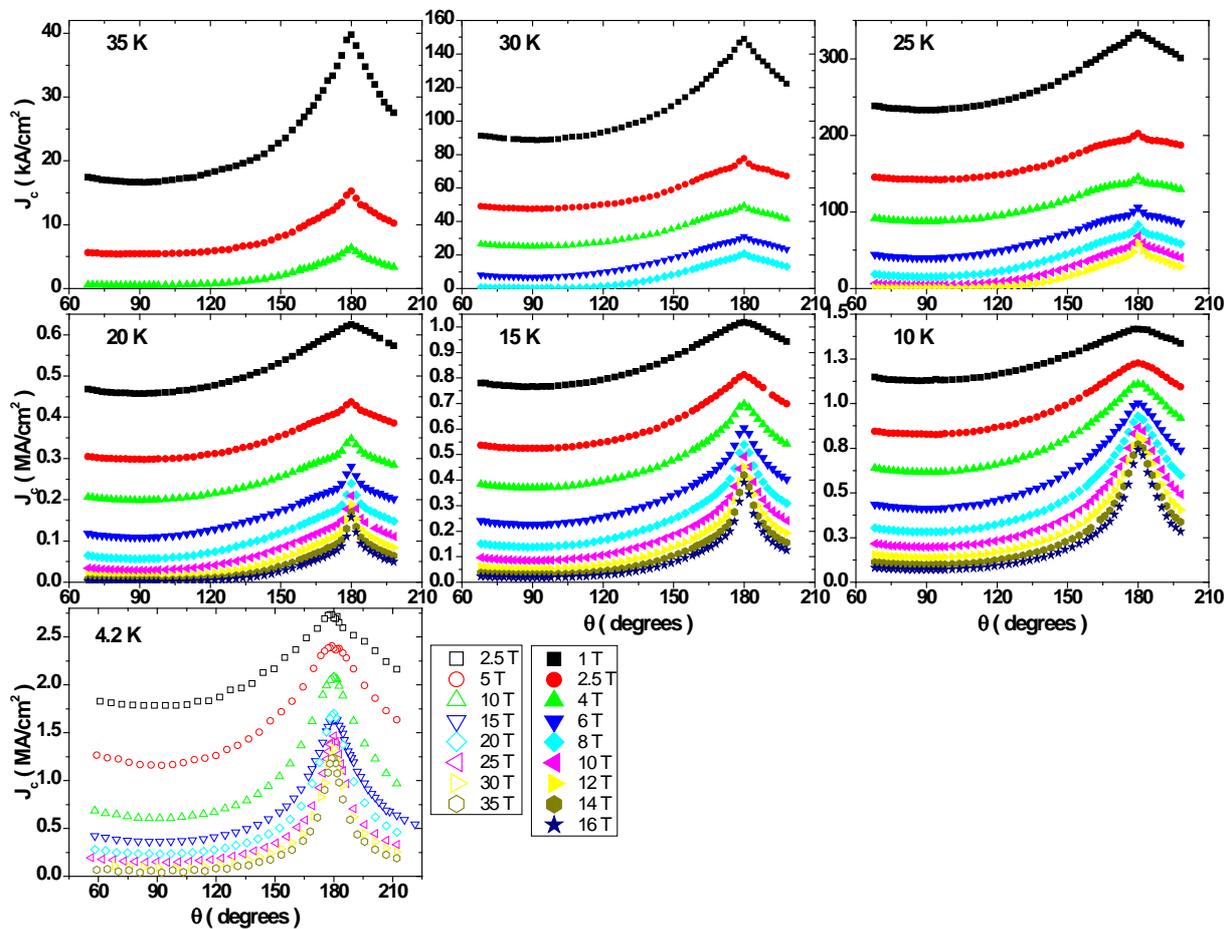

**Figure S2** Angular dependence of $J_c$ measured between 35 and 4.2 K and up to 35 T at 4.2 K, up to 16 T at higher temperature.